\documentclass[journal=nalefd,manuscript=letter, layout=onecolumn]{achemso}

\usepackage[version=3]{mhchem} 
\usepackage{xcolor}
\usepackage{latexsym, amsfonts, amssymb, txfonts, pxfonts, wasysym}
\usepackage{ulem} 

\author{Zhengguang Lu}
\affiliation[NHMFL]
{National High Magnetic Field Laboratory, Tallahassee, FL 32310, USA}
\alsoaffiliation[FSU]
{Physics Department, Florida State University, Tallahassee, FL 32306, USA}

\author{Daniel Rhodes}
\affiliation[Columbia]
{Department of Mechanical Engineering, Columbia University, New York, NY 10027, USA}

\author{Zhipeng Li}
\affiliation[RPI]
{Department of Chemical and Biological Engineering, Rensselaer Polytechnic Institute, Troy, NY 12180, USA}

\author{Dinh Van Tuan}
\affiliation[Rochester1]
{Department of Electrical and Computer Engineering, University of Rochester, Rochester, New York 14627, USA}

\author{Yuxuan Jiang}
\affiliation[NHMFL]
{National High Magnetic Field Laboratory, Tallahassee, FL 32310, USA}

\author{Jonathan Ludwig}
\affiliation[NHMFL]
{National High Magnetic Field Laboratory, Tallahassee, FL 32310, USA}
\alsoaffiliation[FSU]
{Physics Department, Florida State University, Tallahassee, FL 32306, USA}

\author{Zhigang Jiang}
\affiliation[GaTech]
{School of Physics, Georgia Institute of Technology, Atlanta, GA 30332, USA}

\author{Zhen Lian}
\affiliation[RPI]
{Department of Chemical and Biological Engineering, Rensselaer Polytechnic Institute, Troy, NY 12180, USA}

\author{Su-Fei Shi}
\affiliation[RPI]
{Department of Chemical and Biological Engineering, Rensselaer Polytechnic Institute, Troy, NY 12180, USA}

\author{James Hone}
\affiliation[Columbia]
{Department of Mechanical Engineering, Columbia University, New York, NY 10027, USA}

\author{Hanan Dery }
\affiliation[Rochester1]
{Department of Electrical and Computer Engineering, University of Rochester, Rochester, New York 14627, USA}
\alsoaffiliation[Rochester2]
{Department of Physics and Astronomy, University of Rochester, Rochester, New York 14627, USA}
\email{hanan.dery@rochester.edu}

\author{Dmitry Smirnov}
\affiliation[NHMFL]
{National High Magnetic Field Laboratory, Tallahassee, FL 32310, USA}
\email{smirnov@magnet.fsu.edu}

\title{
Magnetic field mixing and splitting of bright and dark excitons in monolayer MoSe$_2$}

\begin{document}

\newpage
\begin{abstract}
\small {
Monolayers of semiconducting transition metal dichalcogenides (TMDCs) with unique spin-valley contrasting properties and remarkably strong excitonic effects continue to be a subject of intense research interests. These model 2D semiconductors feature two fundamental intravalley excitons species - optically accessible ' bright' excitons with anti-parallel spins and optically inactive 'dark' excitons with parallel spins. For applications exploiting radiative recombination of bright excitons or  long lifetime dark excitons, it is essential to understand the radiative character of the exciton ground state and establish the energy separation between the lowest energy bright and dark excitons. Here, we report a direct spectroscopic measure of dark excitons in  monolayer MoSe$_2$ encapsulated in hexagonal boron nitride.
By applying strong in-plane magnetic field, we induce mixing and splitting of bright and dark exciton branches, which enables an accurate spectroscopic determination of their energies.  We confirm the bright character of the exciton ground state separated by a 1.5~meV gap from the higher energy dark exciton state, much smaller compared to the previous theoretical expectations. These findings provide critical information for further improvement of the accurate theoretical description of TMDCs electronic structure.
}
\end{abstract}

\includegraphics[width=14cm]{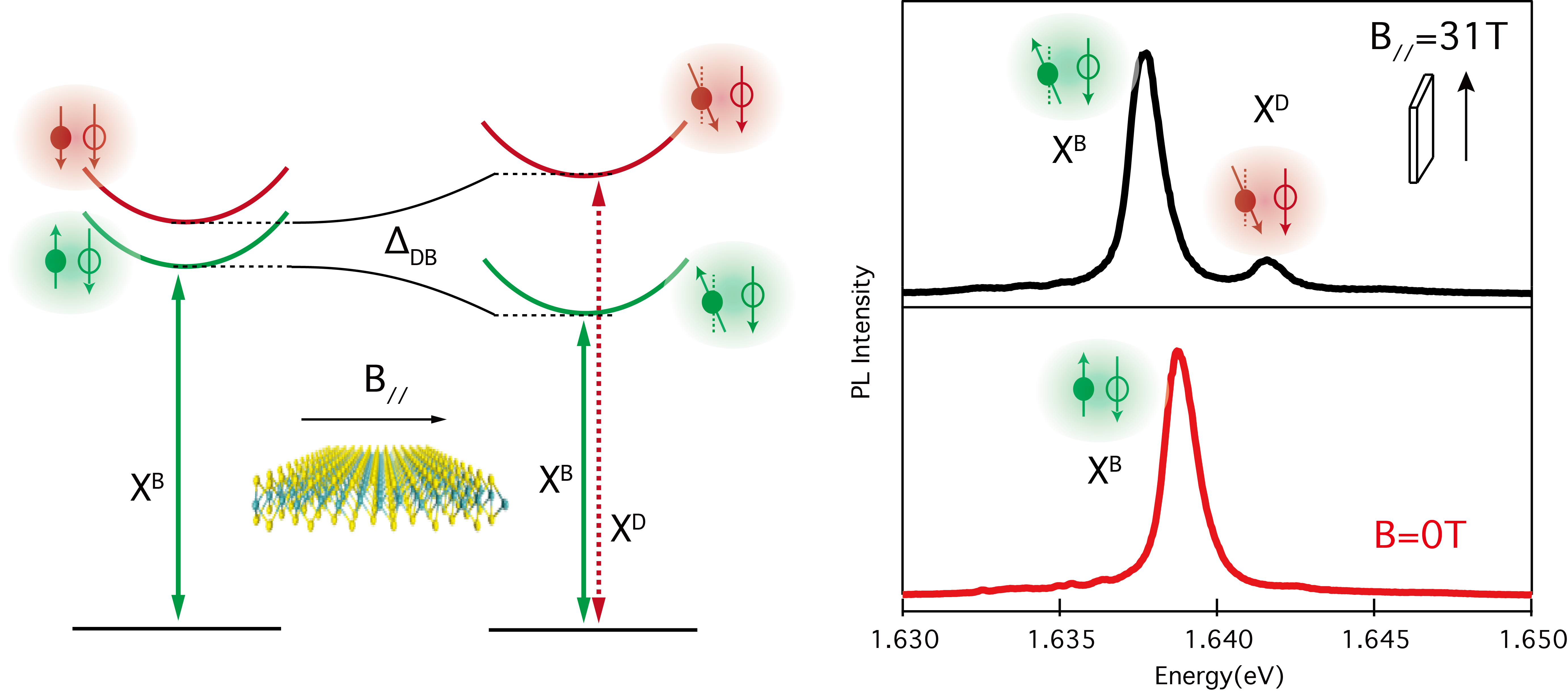}

\clearpage
\newpage

Transition metal dichalcogenides (TMDCs) with the chemical formula MX$_2$,  where M is a group IV-VII transition metal and X is a chalcogen atom (S, Se, Te), have recently emerged as a new class of layered materials that can be exfoliated down to atomically thin layers. Among these materials, the group VI (Mo, W)  TMDC monolayers stand out due to their peculiar  spin-valley coupled  band structure with spin-polarized conduction and valence band states  located at the band edges at the corners of the hexagonal Brillouin zone (K+ and K- valleys)  with a direct gap in the visible or near-infrared region \cite{Xiao_2012, Mak_2010, Splendiani_2010}. Because of time reversal symmetry, the sign of the spin splitting is opposite in K+ and K- valley. This allows one to address the spin-valley degree of freedom with either left or right circularly polarized light \cite{Mak:2012bh, Cao:2012eu} and, together with the strong excitonic effects \cite{Yu-Cui-Xu_2015}, create a unique platform to study both light-matter interactions and spin-valley physics in the real two-dimensional (2D) limit.   

Since allowed interband optical transitions must conserve the spin and the momentum, the optically active or bright exciton involves an electron and  a hole belonging to the same valley 
but with opposite spins (the spin of the hole is opposite to that of the electron residing in the same valence band). Radiative recombination of intarvalley excitons with the electron and hole constituents with same spins is forbidden and such exciton states are called dark. The understanding of the interplay between low energy bright and dark exciton states is of particular importance since these states ultimately define the radiative properties of 2D TMDCs and their relevance for  applications in optoelectronics and valleytronics. In WSe$_2$ and WS$_2$ monolayers, the dark character of the  ground state exciton and the energy separation between dark and higher energy bright excitons are firmly established both experimentally and theoretically \cite{Zhang_NatNano17, Zhou-Park_2017, Wang_PRL17, Molas_2017, Vasconcelos_PRB18, Park-Raschke_NatNano18, Molas_arXiv19}. Meanwhile, the low energy dark excitons in Mo-based compounds have not been probed experimentally so far, and recent theoretical studies have not always been consistent in predicting whether Mo-based TMDCs host dark or bright lowest energy excitons \cite{Palummo_2015, Dery-Song_2015, Echeverry_PRB16, Deilmann_PRB17, Malic_PhysRevMat18}.    

In this letter, we present the first measurement of the dark exciton in monolayer MoSe$_2$. The main obstacles to detecting dark excitons in MoSe$_2$ are the very close separation between the bright and the higher energy dark states and the low population of the dark exciton states \cite{Wang_PRL17}. Due to recent progress in the fabrication of high quality BN encapsulated TMDC monolayers, it has become possible to prepare TMDC monolayers with very narrow exciton lines of about 1~meV FWHM \cite{Ajayi_2017}.  
With the aid of a strong in-plane magnetic field,  we mix bright and dark exciton states, brighten dark excitons, induce further separation of bright and dark exciton states and successfully detect the elusive dark excitons in photoluminescence (PL) and reflectance contrast (RC) spectra.  
These experimental results provide an unambiguous evidence of the bright nature of exciton ground state in MoSe$_2$ as well as the accurate measure of the energy splitting between the lowest energy bright and dark excitons.

Within a simple single particle picture, the radiative character, dark or bright,  of  intravalley excitons in monolayer TMDCs  depends on the ordering of the spin-up and spin-down conduction sub-bands, CB$\uparrow$ and CB$\downarrow$,  separated by a relatively small single-particle gap $\Delta_{CB}$ of the order of tens of meV \cite{Kosmider_SO-CB_2013}.
For the valence states, only the higher energy spin-subband VB$\uparrow$ needs to be considered because of the large, hundreds of meV,  spin splitting in the valence band \cite{Zhu_SO-VB_2011}.
The relative ordering of lowest  energy conduction subbands in MX$_2$ monolayers, which are formed mostly by  $d_{z^2}$ orbitals of transition metal atoms  \cite{Liu_2013}, is different for Mo- and W-based TMDCs. In molybdenum compounds, the lowest conduction band and the the highest valence band share the same spin projection, so the lowest energy exciton state is expected to be optically active.  However, the actual energy separation between the lowest energy bright and dark excitons states,  $\Delta_{DB}$, 
has additional contributions from the short-range electron-hole exchange interaction as well as enhanced binding energy of dark excitons due to their heavier mass compared with bright excitons (see the theory section in the Supporting Information) \cite{Echeverry_PRB16, Deilmann_PRB17}. 

In order to access the dark exciton states, we apply an in-plane magnetic field $B_\parallel$ ($B_x$,$B_y$)  to separate more dark ($X^D$) and bright ($X^B$) excitons and  brighten  otherwise unaccessible dark excitons. This is illustrated in Fig.\ref{fgr:1}a where we show a cartoon of the dispersion relation of excitons as a function of the exciton center-of-mass wavevector ($Q$). Only excitons located within the photon dispersion cone are allowed to recombine radiatively. The in-plane magnetic field does not couple to the orbital motion of the electrons and can effectively perturb the spin magnetic momenta of band electrons only. The effect of B$_{//}$ on the spin orientation of valence band electrons is negligible because of the large separation of spin-orbit split valence bands. Thus, the system can be approximated by a two-level Hamiltonian for bright and dark exciton states (see the theory section in the Supporting Information): 

\begin{figure}[t]
\centering
\includegraphics[width=15cm]{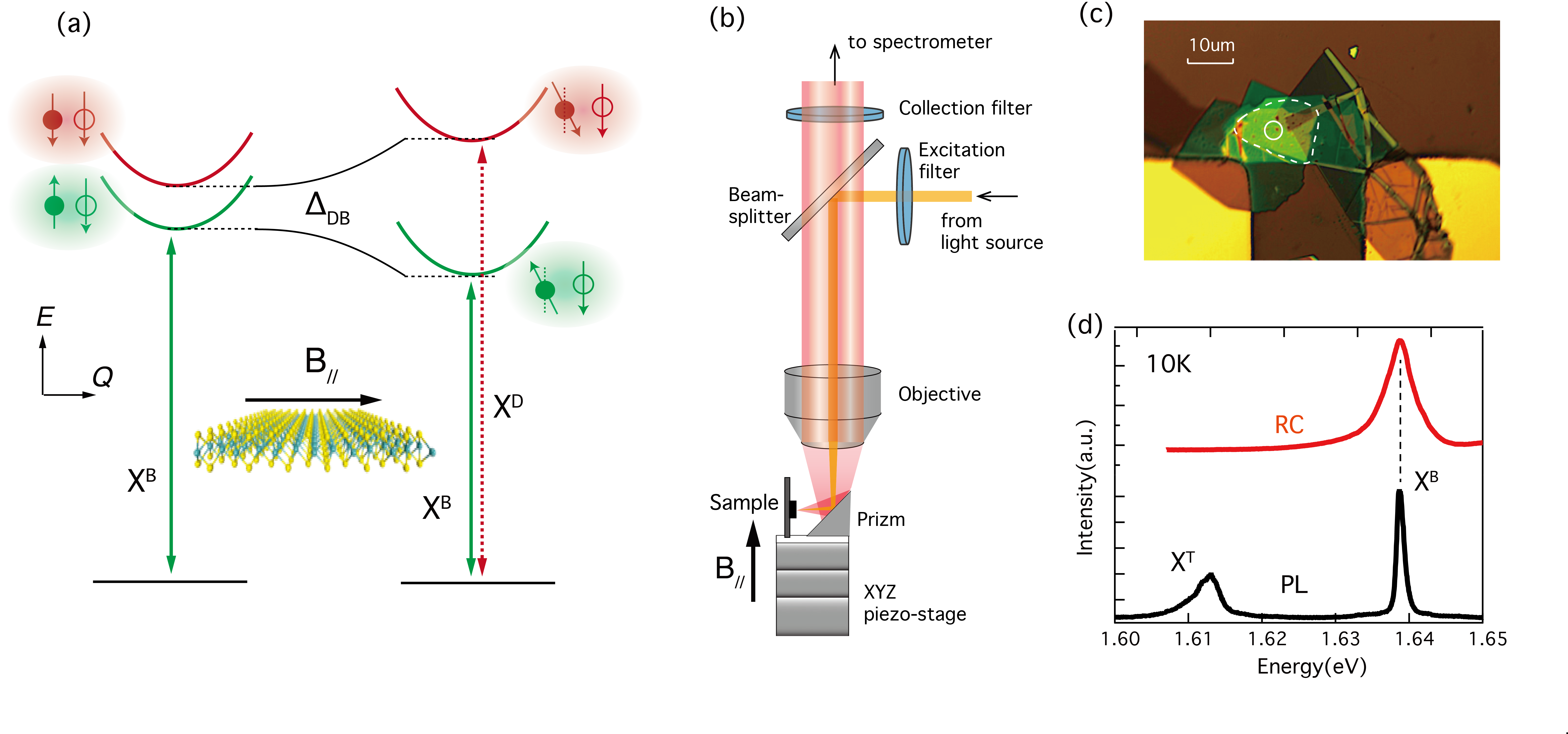}
\captionsetup{justification=justified, singlelinecheck=false}
\caption{
(a) Electronic dispersion of spin-allowed bright ($X^B$, depicted in green cartoons) and spin-forbidden dark ($X^D$, depicted in red cartoons) intravalley excitons in monolayer MoSe$_2$. 
The in-plane magnetic field mixes spin components of CB electrons, which transfer some oscillator strength from bright excitons to dark ones and further separate them in energy. In the experiment, this mechanism can be probed via magneto-optical spectroscopy in the Voigt  configuration.     
(b) Schematics of the experimental setup for magneto-PL and magneto-reflectance measurements in the Voigt geometry.
(c) The microscope image of the hBN encapsulated MoSe$_2$  device. The dashed line encloses the  MoSe$_2$ monolayer. The excitation light spot is approximately indicated by the center circle.
(d) PL and broadband RC spectra measured at 10~K. The bright neutral exciton ($X^B$) is observed in both PL and RC spectra. The PL emission at 26~meV below the neutral exciton peak is attributed to the trion (X$^T$) emission.
}
\label{fgr:1}
\end{figure}

\begin{equation}
{ H }_{ eff }=\begin{pmatrix} { E }_{ B } & \frac { 1 }{ 2 } { g }_{ cb }{ \mu  }_{ B }  { B }_\parallel   \\ \frac { 1 }{ 2 } { g }_{ cb }{ \mu  }_{ B } { B }_\parallel  & { E }_{ B }+{ \Delta  }_{ DB } \end{pmatrix}
\end{equation}
where $E_B$ and $E_D = E_B + \Delta_{DB}$  refer to the energies of the bright and dark states at zero magnetic field, the off-diagonal terms introduce their field-induced coupling,  g$_{cb}$ is the in-plane conduction band $g$-factor and $\mu_B$ is the Bohr magneton. $B_\parallel$ shifts the energies of bright and dark excitons in opposite directions:

\begin{equation}
E _{ B/D }(B_\parallel) ={ E }_{ B }+\frac { { \Delta  }_{ DB } }{ 2 }  \mp  \sqrt { { \left( \frac { { \Delta  }_{ DB } }{ 2 }  \right)  }^{ 2 }+{ \left( \frac { { g }_{ cb }{ \mu  }_{ B } }{ 2 }  \right)  }^{ 2 }{ B_\parallel}^{ 2 } }  
\label{Eq:B-D}
\end{equation}
where the plus /minus sign applies to the bright/dark exciton branches.  $B_\parallel$ also provides small but finite optical oscillator strength, proportional to ${B_\parallel}^{2}$, for dark excitons.  

Magnetic brightening of dark excitons has been observed in tungsten-based TMDCs \cite{Zhang_NatNano17, Molas_2017}, where the gap between bright and dark excitons is relatively large, 
$\sim$50 meV. However, the actual field induced splitting between the bright and dark excitons has not been detected. From  Eq.~\ref{Eq:B-D} we expect the change of exciton energies at the highest applied magnetic fields (31~T) to only be about (55-75)~$\mu$eV for WS$_2$ or WSe$_2$, almost impossible to detect by standard methods. In contrast, since the field induced splitting is inversely proportional to the zero-field $\Delta_{DB}$ (in the weak-field approximation), it is expected to be much stronger in MoSe$_2$, eventually reaching the meV scale  at 20-30~T.

Fabricating high quality MoSe$_2$ monolayers with narrow PL linewidth  of about a 1~meV  is a crucial factor for detecting the exciton fine structure. 
Our samples were prepared by mechanical exfoliation of CVT or flux grown bulk MoSe$_2$ single crystals, then transferred and encapsulated in hexagonal boron nitride (hBN). A piece of few-layer graphene was used as the contact electrode of the single layer MoSe$_2$ and another piece was used as the transparent top-gate electrode on the top layer BN.
The typical size of the monolayer MoSe$_2$ is about 10~$\times$10$\mu$$m^2$ (Fig.\ref{fgr:1}c). 
Our hBN encapsulated samples exhibit very narrow exciton emission PL peaks (with FWHM as low as $\sim$1.1~meV at 10~K ) approaching the intrinsic limit \cite{Ajayi_2017}.

Low temperature magneto-spectroscopy measurements were carried out in the Voigt geometry (light wavevector perpendicular to $B$) with custom micro-spectroscopy setups equipped with a nonmagnetic objective and a three-axis piezostage for fine \textit{in-situ} positioning as illustrated in Fig.~\ref{fgr:1}b. The PL measurements were performed with a 31~T resistive magnet  using excitation by a continuous-wave laser with a photon energy 2.41~eV (514~nm wavelength). 
To measure the reflectance contrast (RC), $\Delta R / R = (R_{sample} - R_{substrate})/R_{substrate}$,  
we employed a similar setup coupled to a 17.5~T superconducting magnet and used a Hg arc lamp with a combination of appropriate bandpath filters to define the spectral region of interest. The excitation light was focused by an objective (NA=0.5) to a spot size of $\sim$2~$\mu$m. The excitation power delivered to the sample was 1~$\mu$W or less to avoid sample heating and minimize light-induced doping. The PL emission or reflected light was collected by the same objective and detected by a spectrometer equipped with a  TE cooled CCD camera. The presented spectra were collected at 10~K and recorded with a spectral resolution of 0.11~meV.
Characteristic zero field spectra of the monolayer MoSe$_2$ are shown Fig.\ref{fgr:1}d. In accordance with prior studies, we observe strong and narrow PL emission line for neutral bright exciton (X$^B$) at 1.6388~eV and PL signal from charged excitons or trions (X$^T$) at 1.613~eV.  The RC spectrum shows only one pronounced peak associated with X$^B$. 

\begin{figure}[t]
\centering
\includegraphics[width=14cm]{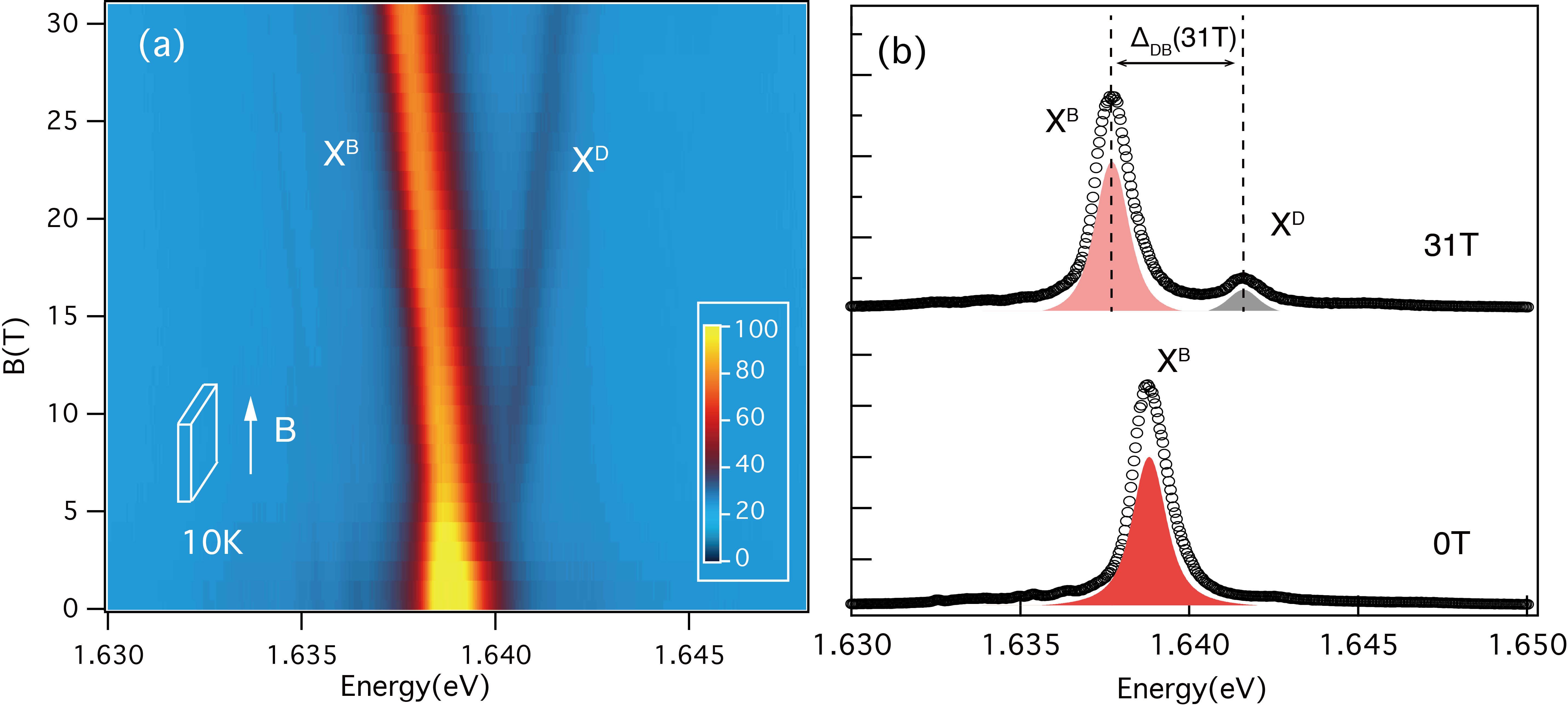}
\captionsetup{justification=justified, singlelinecheck=false}
\caption{
(a) Color map of the evolution of PL spectral intensity as a function of the in-plane magnetic field. The spectra were measured on the hBN encapsulated monolayer MoSe$_2$ in the intrinsic regime. 
(b) Comparison of the PL spectra (open circles) measured at zero magnetic field and at $B_\parallel$ = 31~T. The PL emission energies of X$^B(0T)$=1.6388~eV, X$^B(31T)$=1.6377~eV,  X$^D(31T)$=1.6416~eV, were determined from best peak fits (colored areas).
}
\label{fgr:3}
\end{figure}

To show the effect of the in-plane magnetic field on exciton emission, we plot in Fig.\ref{fgr:3}a  the 2D color map of PL intensity as a function of magnetic field from 0~T  to 31~T. As $B_\parallel$ increases, the  X$^B$ is redshifted and an additional peak, labelled as X$^D$,  emerges on the higher energy of X$^B$  side exhibiting a blueshift. 
At the highest magnetic field  of 31~T,  X$^B$ shifts by 1.1~meV and  the separation between two peaks, X$^B$ and X$^D$,  reaches approximately 4~meV. It is instructive to note that these observations provide an explanation for the results from our earlier attempts to detect field induced bright-dark excitons splitting in the PL spectra from  MoSe$_2$ monolayers on SiO$_2$/Si substrates. Because the exciton PL line was relatively broad, FWHM=7.3~meV, the $\Delta_{DB}$ spitting could not be resolved spectroscopically even at $B_\parallel$=31~T.  However, as illustrated in Supplementary Fig. S1, the field dependent behavior of the exciton PL peak measured on Si/SiO$_2$ MoSe$_2$ (0.8~meV red shift and 3~meV  broadening)  agrees well with the the actual splitting of the bright and dark exciton resolved in PL and RC spectra from hBN encapsulated MoSe$_2$.

\begin{figure}[t]
\centering
\includegraphics[width=14cm]{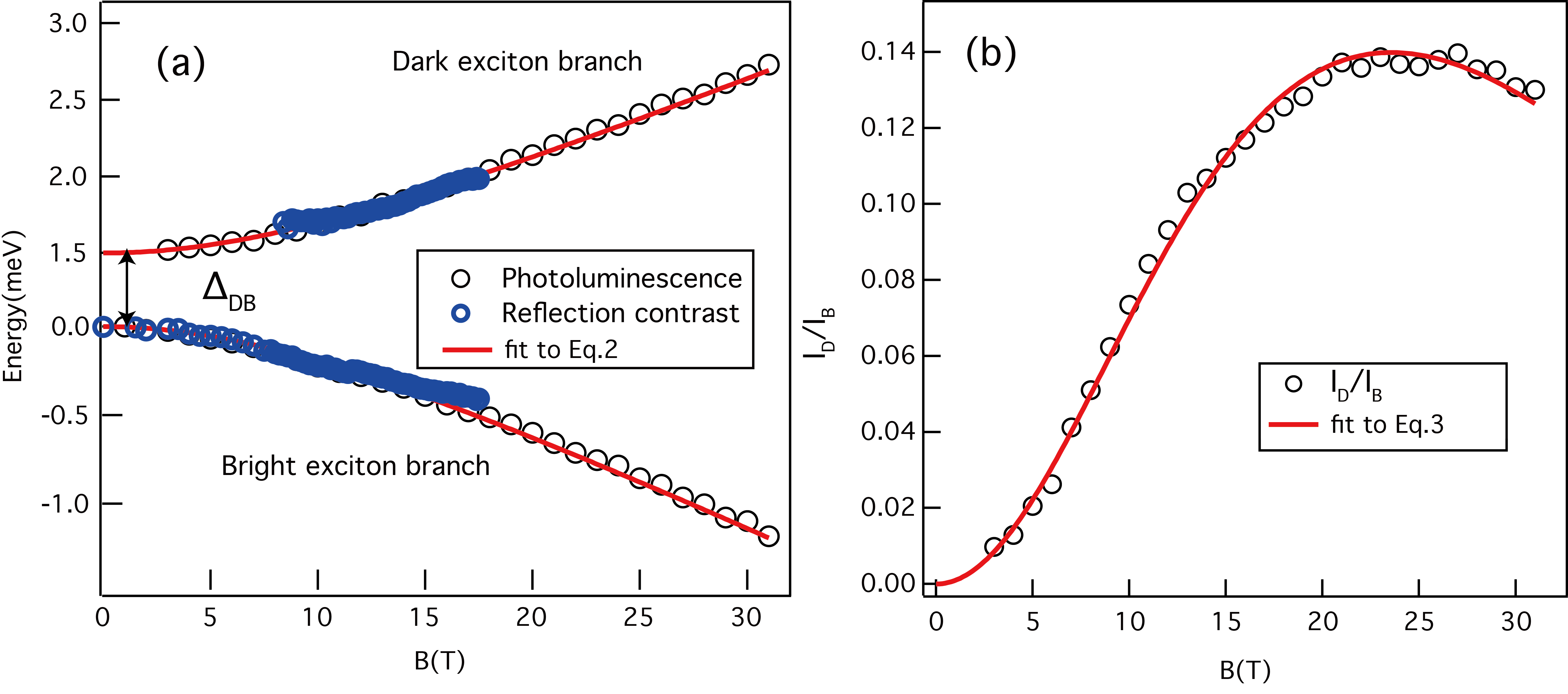}
\captionsetup{justification=justified, singlelinecheck=false}
\caption{
(a) Energy of dark and bright excitons as a function of $B_\parallel$ extracted from PL and RC spectra. Solid lines are fits to Eq.\ref{Eq:B-D}. 
(b) Relative intensity of dark and bright excitons, $I_D / I_B (B_\parallel)$. Solid line represents the fit to Eq.\ref{Eq:Int-T}.
}
\label{fgr:4}
\end{figure}

Since the observed behavior reflects very well the anticipated brightening of the dark exciton and field induced energy shift of bright and dark excitons in opposite directions,  the emerging high energy peak is attributed to the magnetically brightened dark exciton. 
For a quantitative analysis of PL evolution with $B_\parallel$, we use a pseudo-Voigt line shape to fit the spectra. 
The resulting overall change of X$^B$ and X$^D$ energies is plotted in  Fig.\ref{fgr:4}a.  To ensure the reliability and consistency of the experiment and analysis,  we also plot the exciton energies extracted from the RC spectra measured in fields up to 17.5~T (detailed comparison of PL and RC spectra is presented in Supplementary Fig.S1).  
Both data sets consistently overlap, ensuring the validity of the PL results at higher magnetic fields.

As expected from the two band model, the energy of the bright (dark) exciton first exhibits a quadratic increase (decrease) followed by a linear dependence in the high field limit. 
Despite its simplicity, the model provides a facile means for accurate measure of the dark exciton energy at zero magnetic field and the  $g$-factor.
By fitting the measured $E_B(B_\parallel)$ and $E_D(B_\parallel)$ (see Fig.3a) we extract a g-factor of 2.0 $\pm$ 0.1. This confirms the  assumption that  $B_\parallel$ acts through mixing spins of the conduction band electrons, and the orbital or valence band contributions are negligible. 
For the zero-field dark exciton, we obtain $E_D$=1.6403~eV, which places it above the bright state by just  $\Delta_{DB}$=1.5~meV. 
This result represents the first unequivocal experimental observation of dark excitons in monolayer MoSe$_2$ and provides a clear evidence that the exciton ground state in this system is optically bright.

The energy splitting between the bright and dark excitons at the light cone, $\Delta_{\text{DB}} =   \Delta_{CB}  +  \Delta_x + \Delta_M$, has contributions from spin-orbit coupling energy splitting of the conduction band ($\Delta_{CB}$),  short range electron-hole exchange ($\Delta_x$), and the fact that dark and bright excitons have different masses and therefore different binding energies ($\Delta_M$). To compare the experimental value of $\Delta_{\text{BD}}$ with theory, one has to accurately calculate the values of $\Delta_{CB}$, $\Delta_x$ and $\Delta_M$. 
To date, such a comparison is ineffective because the variation in reported ab-initio calculations of these parameters is often larger than 10~meV \cite{Echeverry_PRB16, Kormanyos_2DMater15, Deilmann_PRB17,Zhang_NatNano17}, i.e. larger than our measured value of $\Delta_{\text{DB}}$ in monolayer MoSe$_2$. 
Nonetheless, we can still make few consistent observations. Firstly, we note that $\Delta_x < 0$ due to the repulsive nature of the short-range exchange \cite{Echeverry_PRB16,Deilmann_PRB17,Zhang_NatNano17}, and $\Delta_M < 0$ because dark excitons are heavier than bright ones (dark excitons involve the conduction band with heavier effective mass). Accordingly, to offset the negative value contributions of $\Delta_x+\Delta_M$, the value of $\Delta_{CB}$ should be positive in monolayer MoSe$_2$ in order to yield  $\Delta_{\text{DB}}=1.5$~meV. Similarly,  assuming that $\Delta_x+\Delta_M$ has similar magnitude in Mo- and W-based compounds, the value of $\Delta_{CB}$ should be negative in WSe$_2$ and WS$_2$ monolayers in order to yield $\Delta_{\text{DB}}= -40$~meV and $-55$~meV, respectively \cite{Wang_PRL17}.  
Next, following the procedure described in the Supporting Information, we evaluate the binding energies of bright and dark excitons and obtain $\Delta_M = -8$~meV in MoSe$_2$ and $\sim$-16~meV in WSe$_2$. Assuming  $\Delta_x \sim -10$~meV  \cite{Echeverry_PRB16, Zhang_NatNano17}, we can further infer that $\Delta_{CB}$ is in the ballpark of $+20$~meV in  monolayer MoSe$_2$ and $-20$~meV in monolayer WSe$_2$. The latter is consistent with the measured value reported in Ref.~\citenum{Zhang_NatNano17}. 
We note that in addition to the contribution from spin-orbit coupling, the value of $\Delta_{CB}$ is affected from band-gap renormalization in electron-doped samples due to many-body exchange interactions \cite{Scharf_JPCM19, VanTuan_PRB19}. The experimental results and the analysis we have provided in this paper deal with intrinsic (un-doped) monolayers.

We now discuss the relative intensities of the PL emission of dark and bright excitons, $I_D/I_B$ that we plot as a function of $B_\parallel$ in  Fig.\ref{fgr:4}b.  At low magnetic fields, the dark exciton emission grows quadratically with  $B_\parallel$,  which is in agreement with the predictions of the two-band model, $I_D \propto I_B B_\parallel^2$, and previous observations in W-based monolayer TMDCs \cite{Zhang_NatNano17, Molas_2017, Molas_arXiv19}.  However, we find that in MoSe$_2$, the ratio of PL intensities $I_D/I_B$  deviates from the simple quadratic dependence at magnetic fields above 7-10~T, saturates at higher fields  and starts decreasing above 25~T (Fig.\ref{fgr:4}b). To understand this behavior,  we have to consider changes in the thermal population of $X^D$. 
Since the bright exciton is the ground excitonic state and the separation between bright and dark states increases with $B_\parallel$,  the population of the high-lying dark state becomes thermally suppressed as magnetic field increases. Assuming thermal equilibrium occupation numbers, we modify the expression for the relative intensity  by adding a Boltzmann factor: 

\begin{equation}
 I _ D \propto I_B { B_\parallel }^{ 2 }{ e }^{ \frac { -{ \Delta }_{DB}(B_\parallel) } { { k }_{ B }T }  },
\label{Eq:Int-T}
\end{equation}
which results in excellent fit to the experimental data (Fig.\ref{fgr:4}b).

\begin{figure}[t]
\centering
\includegraphics[width=17cm]{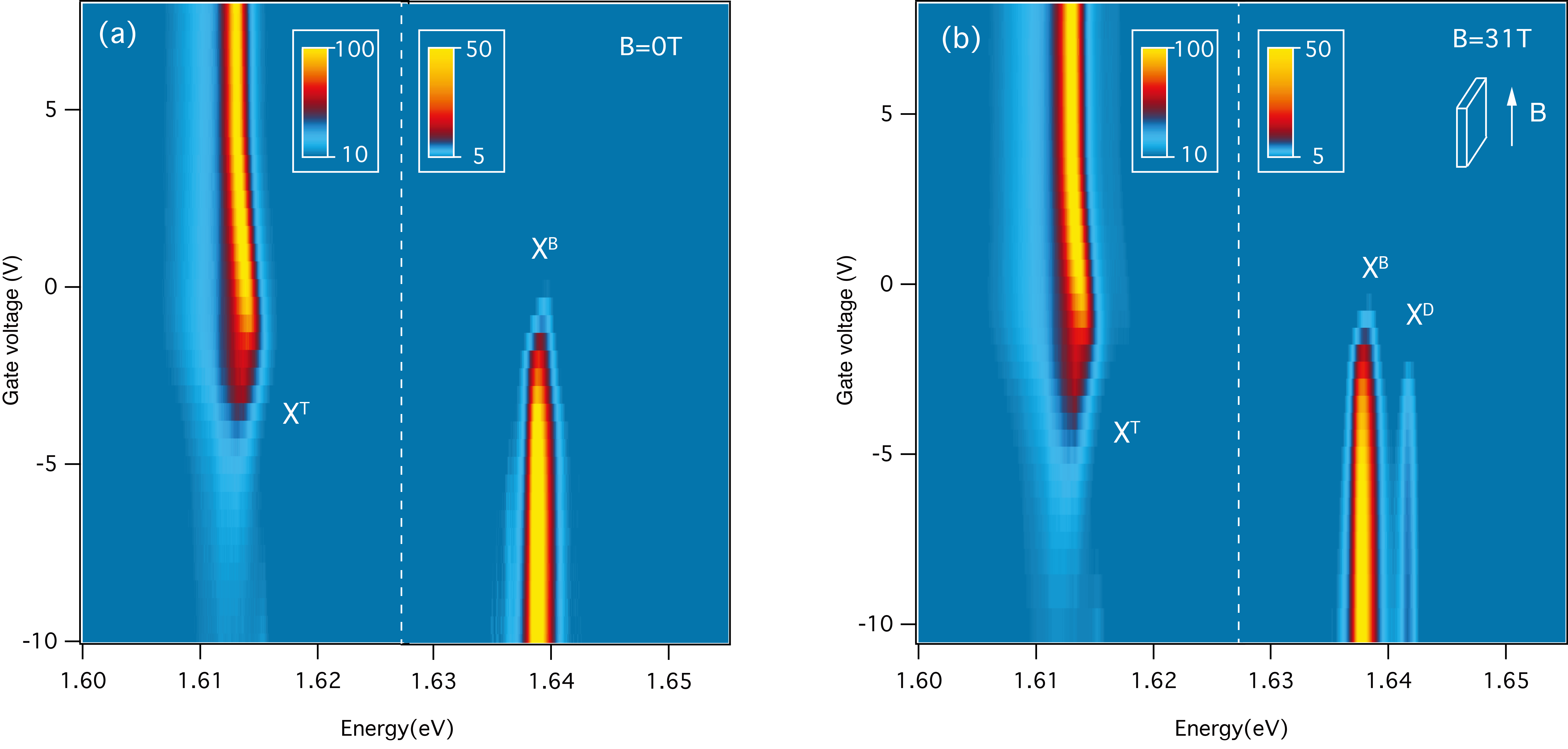}
\captionsetup{justification=justified, singlelinecheck=false}
\caption{
Gate dependence map of the PL intensity at 0 T (a) and $B_{\parallel}$ = 31~T (b) and at $T$=10~K.  The intensity is plotted on a linear scale on the left sub-panels at energies below 1.6275~eV (indicated by dashed white line).  The log scale with smaller range is used to enhance visibility on the right sub-panels (above 1.6275~eV).  $X^B$ and $X^D$ represent bright and dark excitons. $X^T$ is the negatively charged exciton (trion). 
}
\label{fgr:5}
\end{figure}

Finally, we briefly discuss another possible mechanism, the  Bychkov-Rashba effect \cite{Bychkov-Rashba_1984}, that may flip the spin of the conduction band electrons and contribute to brightening of the dark exciton.
The Bychkov-Rashba type coupling between dark and bright intraband excitons would require breaking the mirror  symmetry in respect to the monolayer plane, which could be induced, for example, by out-of-plane electric field  generated by a gate or by unintentional, impurity related charge of the substrate \cite{Dery-Song_2015, Slobodeniuk_2016}.  To  examine this possibility, we  perform gate dependent PL measurements. 
Figure \ref{fgr:5}a shows a color map of the PL spectrum of the device at zero magnetic field as a function of the top gate voltage. In the whole wide range of applied  gate voltages, only two PL peaks are observed related to neutral bright exciton emission ($X^B$) and charged exciton emission ($X^T$). No signatures of the PL emission on the higher energy side of  $X^B$ can be detected in the spectra acquired at $B=$ 0~T. 
The gate dependent PL measured at $B_\parallel=$31~T (Fig.\ref{fgr:5}b) clearly reveals a signal associated with the emission from dark excitons, $X^D$,  situated at approximately 4~meV above $X^B$.  The energy separation and relative intensities are nearly independent on the gate voltage, 
and, consistent with theoretical estimations, indicates that the electric field does not contribute to the brightening of dark excitons.  
Even at relatively high electric field of 1~V/nm, the resulting radiative rate for dark excitons is expected to be four orders of magnitude smaller compared to the effect of 
of an applied in-plane magnetic field of 30~T \cite{Slobodeniuk_2016}.     
The negligible role of the Rashba-induced mixing between dark and bright excitons in our experiment is corroborated by the fact that this interaction is proportional to the exciton wavevector $Q$ (see the Supporting Information), and therefore its effect vanishes within the light cone wherein $Q \rightarrow 0$.

In conclusion, we have observed magnetic field induced brightening of the dark exciton and splitting of bright and dark neutral excitons in high quality monolayer h-BN encapsulated MoSe$_2$. 
The superior sample quality and the use of a strong in-plane magnetic field enable a direct spectroscopic probe of dark excitons.  
We have confirmed that the lowest energy exciton state is optically active.  The energy separation between the ground state bright exciton and higher energy dark exciton is determined to be 1.5~meV.
This result implies that the energy splitting in the conduction band is a few tens of meV in order to offset the reduced binding energy of bright excitons due to the repulsive short-range electron-hole exchange as well as their lighter mass compared with dark excitons. Importantly,  our findings demonstrate that the bright and dark exciton branches are nearly degenerate, and this result may be intimately related to the enhanced valley depolarization in  monolayer MoSe$_2$ \cite{Wang-Toulouse_APL2015, Kioseoglou_SciRep2016, Tornatzky_PRL2018}.
\\
\\
ACKNOWLEDGEMENTS.
\\
This work was primarily supported by the  Department of Energy, Basic Energy Sciences, under Contract No. DE- FG02-07ER46451.
The MoSe$_2$ crystal growth and the sample fabrication at Columbia University were supported by 
by the NSF MRSEC program through Columbia in the Center for Precision Assembly of Superstratic and Superatomic Solids (DMR-1420634).
The device fabrication at Rensselaer Polytechnic Institute (RPI)  was supported by the NY State Empire State Development's Division of Science, Technology, and Innovation (NYSTAR) through Focus Center-NYÐRPI Contract C150117. S.-F. Shi acknowledges support by AFOSR through Grant FA9550-18-1-0312.
The experiments were performed at the National High Magnetic Field Laboratory (NHMFL), which is supported by NSF Cooperative Agreement DMR-1157490, DMR-1644779 and the State of Florida. 
The work at the University of Rochester was supported by the Department of Energy, Basic Energy Sciences, under Contract No. DE-SC0014349.

\end{document}